# Dynamic Response in Fe doped $La_{0.65}Ca_{0.35}Mn_{1-y}Fe_yO_3$: Rare-earth Manganites.

Wiqar Hussain Shah[1,2], S.K. Hasanain[3]

[1]Department of Physics, College of Science, King Faisal University,
Hofuf, 31982, Saudi Arabia

[2]Department of Physics, Federal Urdu University,
Islamabad, PAKISTAN

[3]Department of Physics, Quaid-i-Azam University, Islamabad, PAKISTAN

**Abstract:**

We have studied the dynamic response of Fe doped manganites with ac susceptibility measurements on bulk $La_{0.65}Ca_{0.35}Mn_{1-y}Fe_yO_3$ with $0.01<y<0.10$ as functions of temperature and dc field. It is observed that the in phase part goes through a maximum that is removed on the application of moderate dc field. DC fields suppress both the components with the strongest effects being at or close to $T_c$. Conduction and ferromagnetism has been consistently suppressed by Fe doping. Increased spin disorder and decrease of $T_c$ with increasing *Fe* content are evident. The variations in the critical temperature $T_c$ and magnetic moment show a rapid change at about 4-5% *Fe*. The effect of *Fe* is seen to be consistent with the disruption of the *Mn-Mn* exchange possibly due to the formation of magnetic clusters. There is clear correlation between the structures in the resistivity and the in and out of phase parts of the susceptibility. With increasing Fe concentration the out of phase part ($\chi^{//}$) peak shift to $T<T_{1/2}$, but it stays consistently at the temperature where the resistivity changes slow down. The shoulder in $\chi^{//}$ disappears above 4% Fe concentration. The peak of $\chi^{//}$ moves to 8 or 10 K higher temperature on the application of a dc field, for 3 & 4 %Fe samples. Consistently two peaks are found in the ratios of $\chi^{//}/\chi^{/}$. We are seeing correspondence of dissipation with the change in R. We also see increasing low temperature dissipation in more strongly Fe doped samples i.e. increasing the Fe, leads to increased spin disorder and dissipation at low temperature. The effect of the dc field is discussed in terms of the suppression of spin fluctuations close to $T_c$ and the changes in coercivity for lower temperatures.

71.30.+h; 75.30.Vn; 75.30.Kz; 75.60.Lr; 72.60.+g

# I.    INTRODUCTION:

During the last few years, perovskite manganites have attracted broad research interest, because of their unusual magnetic and transport properties. The magnetic and resistive behavior in the colossal magnetoresistance compounds [1,2] is well known to be a sensitive function of the lattice strain produced for example by doping on various sites[3,4,5]. The effects of substituting on the *Mn* sites by other transition elements is further known [6] to affect the properties due to the changes produced in the average electron concentration and the shifts in the positions of the $e_g$ and $t_g$ sub-bands. A number of studies [6-10] have been conducted on the effects of replacement of *Mn* by various transition elements, which have included both global and local measurements such as Mossbauer Spectroscopy. The case of *Fe* is particularly interesting because of the large extent to which it can replace *Mn*. Ahn et al [6] have shown that the observed decrease in $T_c$ was consistent with a weakening of the double exchange (DE) on introduction of the *Fe*. It is apparent from various studies that the *Fe* atom substitutes for *Mn* in the +3 state and does not take part in the DE thereby leading to a lowering of $T_c$. Ogale et al [7] showed that their *Fe* doped compounds showed a marked decrease in $T_c$ at 4% Fe doping. They related it to the average *Fe-Fe* separation approaching the size of the charge carriers (polarons) at this concentration. It has also been shown [10, 11] that *Fe* and *Co* doping leads to the formation of locally anti-ferromagnetically coupled spins or clusters with localized spin excitations. Several recent studies including those by neutron scattering have shown that the onset of the para to ferro transition is accompanied by a divergence of the correlation length, signaling the growth of the infinite correlation and an inhomogeneous [12] type of ferromagnetic transition. It can therefore be expected that the effect of the *Fe* doping, which results in the formation of clusters and suppression of conductivity would also be reflected in the effective activation energies for hopping transport [13], both above and below the temperature corresponding to the resistivity maximum $T_p$. As shown by various studies on CMR compounds [14-17] the resistive behavior both above and below $T_p$ can be described in terms of an activated conductivity, albeit with a magnetization dependent activation barrier for $T<T_p$. We have conducted a study of the magnetic and transport effects in a series of *Fe* doped *CMR* compounds of compositions *$La_{0.65}Ca_{0.35}Mn_{1-y}Fe_yO_3$*, with y varying from $0 \leq y \leq 0.1$, and determined the changes in the susceptibility. Our studies are particularly focussed on the effects at low fields, H<1100 Oe, as compared to previous studies. This field range is also of interest from the point of view of potential applications. The doping mechanism for changing $Mn^{+3}/Mn^{+4}$ ratio in CMR materials is quite similar to that as followed for the High $T_c$ superconductors ($HT_cS$) to control effective Cu vacancy. Similar to $HT_cS$ compounds, both Cu and Mn can be substituted partially by other 3d metals like Co, Ni, Fe and Zn. In case of CMR compounds, Mn (3d) metal site substitution was also carried out with Fe and the results were explained in terms of $Mn^{+3}/Mn^{+4}$ ratio. $T_p$ decreases monotonically with increasing Fe concentration and disappeared for 18% Fe concentration. As expected, for most of the polycrystalline bulk materials, $T_p$ is not sharp for higher y, and paramagnetic (FM) to ferromagnetic (FM) transition takes place over a broad temperature range.

The distortion of the Mn-O-Mn bond angle increases with y so that FM DE interaction weakens, as we expected. At the transition temperature $T_p$, the CMR is highly sensitive to x, the doping concentration and $\delta$, the oxygen deficiency in $R_{1-x}A_xMnO_{3\pm\delta}$ (R=La, Pr, Nd etc and A= Ca, Sr etc) samples. By changing x with fixed $\delta$ or vice versa, one essentially changes the $Mn^{+3}/Mn^{+4}$ ratio in these compounds. There exist a clear relation between $T_p$ and the amount of $Mn^{+4}$ ions. There is clear evidence that the ferromagnetic order does not develop uniformly in the CMR compounds whether optimally doped [18] (with e.g. 33% calcium) or otherwise [19] (e.g. $La_{0.7}Pb_{0.3}MnO_3$), but rather proceeds through the growth of clusters. Close to $T_c$ there is clear evidence [18] for the growth of an extended cluster which undergoes a critical slowing down while smaller clusters grow and are later increasingly incorporated into the extended one. This has been shown to manifest itself in the occurrence of two different time scales for magnetic response [18]. Ibarra et al [20] have shed light on the field dependence of this cluster growth process and the growth of the correlation length. It has been shown [21] theoretically that in general, away from half filling of the $e_g$ sub-bands, there is a tendency for the development of small segregated charge ordered regions or clusters.

Using the technique of ac susceptibility, we have studied the development of spin correlations and their magnetic field dependence in a composition (0.01<y<0.1). We focus particularly on the magnetic transition and the effects related to short range ordering. In general, the out of phase part of the ac response [22], which contains much useful information about the loss mechanisms at and away from a transition, has been studied. It is the objective of this work to investigate both the in and out of phase ac response of a magnetic composition, $La_{0.65}Ca_{0.35}Mn_{1-y}Fe_yO_3$ as a function of field, temperature and to a limited frequency (f=311Hz). We relate the temperature and the field dependence of the susceptibilities to the development of magnetic order, the growth or suppression of fluctuations and also relate these observations to the changes in the resistivity.

## II. Experiment:

All samples reported in the present study were synthesized by standard solid state reaction procedure. Stoichiometric compositions of *$La_{0.65}Ca_{0.35}Mn_{1-y}Fe_yO_3$* (y = 0.0 – 0.10) were prepared by mixing the equimolar amounts of *$La_2O_3$*, *$CaCO_3$*, *$Mn_2O_3$* and *$Fe_2O_3$* (having

+99.9% purity). Powder of these oxides and carbonate were mixed thoroughly in acetone and were finely ground in an electric grinder for thirty minutes. After drying, the mixtures were calcined in alumina boats at 1000 °C for 16 hours, then cooled to room temperature, reground and again heated at 1100 °C for 17 hours. Following cooling to room temperature, they were reground and again heated at 1200 °C for 17 hours. After the third heat treatment, the materials were ground to fine powder and were pressed in to pellets of 13mm diameter and 2 mm thickness under a pressure of 5 tons/inch$^2$. These pellets were heated at 1250 °C for 17 hours. After this final heat treatment laboratory X-ray diffraction (XRD) measurements were carried out to confirm that single phase materials had been prepared [6].

The XRD data were collected by step scanning over the angular range of $15° \leq 2\theta \leq 70°$ at a step size of 0.02° and counting time of 3 seconds per step. We do not observe any appreciable change in the lattice parameters with increasing *Fe* concentration. These results are in agreement with previous findings [6].

## III. Results and Discussion:

The electrical resistance was measured by using the standard four-probe technique using an air drying conducting silver paste. Sample temperature was monitored by a calibrated *Rh-Fe* thermometer in the range of 77 – 300 K. The temperature accuracy was 0.05 K. A constant current in the range of 100 μA and 1 mA was supplied by the current source and voltage across the sample was measured with a digital voltmeter.

All the samples studied here showed a metal-semiconductor (insulator) transition with a characteristic peak in the resistivity at the transition temperature $T_p$. Ferromagnetic transition temperatures $T_c$ will be shown to be very close to these insulator-metal transition temperatures. **Fig.1** shows the dependence of resistivity as a function of temperature for the *$La_{0.65}Ca_{0.35}Mn_{1-y}Fe_yO_3$* (y = 0.0 – 0.1) samples. It is clearly seen that transition temperatures are systematically lowered and peak resistivity values increase with the increase in *Fe* concentration. This is consistent with the general trend in these systems where doping with *Fe* tends to result [6, 7, 8, 9] in a weakening of double exchange interaction, thereby lowering $T_c$. This effect is understood to be primarily due to the *Fe* ions being in the +3 state and hence not participating in the double exchange. The resistivity changes on doping are observed by us to be gradual and linear below 4% *Fe* concentration, while above this concentration, changes are very large. Note that the resistance is plotted on a log scale for convenience. We note that the addition of 10% *Fe* leads to a change of almost six orders of magnitude in the peak resistance. Between 8 and 10% concentration, the peak resistance changes by almost 2 orders of magnitude while the critical temperature decreases by 30 K. The very pronounced change in the resistivity for y>8% is very important results which suggests that the disorder exceed a critical threshold beyond 8%. It is possible that the threshold is connected with the percolation paths in the doped materials.

**AC-susceptibility studies**:

AC studies were conducted using a self-made ac probe with a split secondary (astatically wound) and a commercial lock-in amplifier. The sample could be moved between the two secondary coils that enabled a careful cancellation of the background. The ac-probe was placed inside a commercial Oxford instruments cryostat. The uncompensated coil background was temperature dependent but was cancelled to a good accuracy by making measurement in each of the two counter wound coils, and taking their difference [23],

$$V_1 - V_2 = (V_B + Vs_1) - (V_B - Vs_2).$$

Here $V_1$ is the net signal with the sample placed in coil 1 and similarly for $V_2$. $V_B$ is the total coil background and $Vs_1$, $Vs_2$ are the sample signals in coil 1 and 2 respectively. Due to the counter-wound secondary coils the sample signal in each of the two coils is of opposite sign. Phase adjustment was typically made very far from the transition temperature with a very low amplitude signal ($h_{ac}$=1 Oe), and the desired frequency. The $90^o$ out of phase component was identified as the phase where the total signal in the two coils was identical. i.e. with no out of phase response from the sample, $Vs_1^{//} = Vs_2^{//} = 0$ and thus $V_1=V_2 = V_B$. The signal in-phase with the driving field was determined from it. Since phase setting is independent of ac amplitude the field was raised to desired value once the phase has been adjusted. The susceptometer had been tested for linearity with ac field amplitudes and frequency independent behavior in our range of studies. Our studies were conducted in the range $h_{ac}$=5Oe, f= 311Hz and $0 < H_{dc} < 1100$ Oe. DC magnetic fields were obtained from a homemade solenoid magnet with the dc field direction parallel to that of ac field.

Studies with superposed dc fields were conducted during warm up after the sample had been cooled down to 81K in the desired field. Data were taken on somewhat bar shaped samples with masses of 11.2, 14.7, 14.1, 18.1, 10.5, 14.5, 13.2 and 13.8 mgs. respectively. Most of the data shown is for the sample with the smaller mass.

The in-phase part of the susceptibility $\chi'$, for Fe doped compositions are studied in detail. All the composition goes from paramagnetic insulator to ferromagnetic metal at a particular temperature region called the transition region. As we increases the Fe concentration the sharpness in the transition region from paramagnetic to ferromagnetic is decreased.

The moments are also decreased from lower concentration towards higher one. The decrease from the peak to the lowest temperature moment is also decreased. The increase in Fe concentration leads to increased spin disorder and dissipation at low temperature which is the most important result of this study. The resistivity peak matches with the tail of the $\chi'$, for all the composition. The valley of the resistivity matches with the midpoint of $\chi'$ for 0 and 1% Fe concentration.

Typical data for the in phase part of the susceptibility, $\chi'$, is shown in **Fig.2** for $h_{ac}$=5 Oe, f=311Hz and (a) $H_{dc}$=0 Oe, (b) $H_{dc}$=550 Oe. One consistent feature to emerge from the $\chi'$ measurement at H=0 is the presence of the decrease in $\chi$ for T < close to $T_c$ consistent with the dc magnetization behavior. For H=0 there is roughly a 0- 25% decrease in $\chi$ below the maximum. The ac response $\chi'$ was studied in the more truly ferromagnetic $La_{0.65}Ca_{0.35}MnO_3$ and a very similar increase in $\chi'$ was observed with increasing temperature culminating in a very sharp drop at $T_c$. Thus both the dc and ac magnetization show consistent behavior for both the compositions. As well define maximum has to appear in the real part of ac susceptibility at the same temperature to that of the imaginary part. Obviously the PM to FM (pf) transition temperature $T_p$ decreases with increasing *Fe*. This is corroborated by the insulator-to- metal transition temperature see as a large anomaly in the electrical resistivity measurements. The resistivity peak matches with the beginning of the $\chi'$ rise for all the samples. The shoulder in the out of phase part $\chi''$ matches with the resistivity peak in 0 and 1% Fe samples.

## DC field effect on in-Phase part ($\chi'$):

The data for $\chi'$ in the Zero field cooled (ZFC) state as shown in **Fig.3** depicts the typical behavior seen in our range of $h_{ac}$ and $H_{dc}$. Firstly, there is a general decrease of $\chi'$ due to the applied dc field. For a field of 550 Oe there is a decrease of about 27-47% at the $\chi'$ maximum for different Fe concentration. We note that for relatively higher fields, the general response of both the dc and ac susceptibilities is very much the same. The decrease of the in-phase part of $\chi'$ due to an applied dc field at low temperatures (T<<$T_c$) is not surprising. The ac susceptibility $\chi'=(dm'/dh)$ is the so-called "reversible susceptibility" [22] which corresponds to the local slope of the *minor loop* traced by the magnetization due to the cycling of the ac field. In the ferromagnetic state this slope, in general, decreases with increasing applied field. Accordingly, so does the susceptibility. On the other hand the dc slope in the low field region, the so-called differential susceptibility [22] is understood to go through a maximum with increasing fields. Thus the effect of the dc field on the dynamic response can be divided into two parts. In the low temperature part of **Fig.3** the almost temperature independent (normalized) change is due to the change in the local susceptibility. However in the high T region (T>$T_c$) where the maximum effects are observed, these may be ascribed to the spin dynamics within and around the critical region which is very crucial results of this work and this is first time in the history of CMR materials that we did this important investigations. In the critical region, the growth of the correlations implies [24] that the long range spin alignments are in a state of rapid dynamic growth. These correlations are suppressed by the application of static biasing fields, particularly in soft ferromagnetic materials, as shown by Williams et al. [25]. This would lower the ac response very strongly, as is observed. This picture will be seen to be supported by the behavior of the out of phase part $\chi''$.

The other feature observed on application of the dc field is the flattening out of the low T decline of $\chi'$. For example for H=0 the decline down to 81K was seen earlier to be about 0-25%. However for H=550 we find that this decrease in $\chi'$ is reduced to 0-8.88%. For an applied field of 1100 Oe we find that the decline is completely eliminated, and the moments at T=81K and at the previous peak are equal. We have studied this effect in a number of compositions and find similar behavior. The extent of the field induced flattening varies with composition and also the qualitative features are changed to some extent. This field dependence is consistent with the dc behavior discussed earlier. The dc fields, while suppressing the overall response, due to change in the local susceptibility, acts to incorporate the smaller clusters into the extended cluster, suppressing the tendency for anti-ferromagnetic alignments.

## Out of phase response $\chi''$:

As is well understood the out of phase response of a driven system (such as the spin system) is proportional to the losses taking place. These losses are typically large at and around a phase transition. The temperature variation of the out of phase response was studied for ac field amplitudes h=5 Oe, frequency f=311Hz and dc field $0 \leq H \leq 550$ Oe. The out of phase part of the pure composition has a very sharp maximum peak and no shoulder. While y=0.01, 0.02 and 0.03% Fe composition have a shoulder before the main peak. As we increase the Fe concentration the peak become broad, the sharpness becomes less and also the peak value is decreased. The shoulder is disappeared at 4% and above this Fe concentration. Above the 5% Fe concentration there are no remarkable response of $\chi''$ at 5 Oe ac field and f=311Hz. The valley of the resistivity matches with the peak of the $\chi''$ and also the shoulder in $\chi''$ matches with the resistivity peak for 0 and 1% Fe composition. Typical temperature dependence of the out of phase part, $\chi''$, for 1% Fe doped sample is shown in **Fig.4** for f=311Hz. and $h_{ac}$=5 Oe. There is a clear increase below about 260 K which in this case has a shoulder at 257 K; and maximum at 250 K. The response then decreases continuously as T is lowered. A significant part of these data is the appearance of a shoulder. It will be seen that these features are retained at the application of dc field, albeit with somewhat less sharpness. We note that the response does not fall to zero below this broad peak and the shoulder but settles down to a finite value. This low temperature dissipation is in the same region where the in-phase part of $\chi$ reaches a maximum. The continued presence of losses in the low temperature region is consistent with the picture of a gradual process of incorporation of small clusters into the main extended cluster, which generates the low temperature losses, which is very important conclusion. However the first peak or shoulder in $\chi''$, appears (257 K) for 1% Fe composition, somewhat below the maximum in resistance R, and just at the mid of the susceptibility curve shown in **Fig.5**. For a system undergoing a magnetic transition the loss components are, according to the fluctuation dissipation theorem [22], proportional to the fluctuations in the magnetization. This loss component can be associated with the dynamical changes taking place in the critical region. As the dynamical correlations of the spin develop in this region, they are accompanied by energy losses and these are correspondingly reflected in the $\chi''$ behavior. This would correspond to the increase of $\chi''$ at 260K (1%Fe) and to the maximum around the mid-point of the transition, around 250K. The main peak in $\chi''$ however occurs close to $T_c$, the temperature where the magnetization M(T) shows no particular structure except for the slowing down of the changes in M. The resistive curve R(T) however shows a clear inflection point close to this temperature (see Fig.1). This suggests that the mechanisms responsible for the slowing down of the resistivity decrease are magnetic in origin and are associated with the increase in the loss component at 260 K. It is notable that the loss component does not exhibit the global maximum at the middle of the magnetic transition, as is generally seen, but at a temperature (250 K) significantly below it. It would appear from this that the losses continue to grow well up to the point where the magnetization continues to increase rapidly. Similar behavior is also observed in the sample with other Fe composition. We understand this to be originating in the continued rapid growth of the smaller clusters in this region, below $T_c$ itself. It is interesting to note that the μSR [11] and SANS data also shows the continued development of the small clusters down to a temperature about 20 K below $T_c$. At lower temperatures the corresponding μSR amplitude decreases indicating the decrease in the number of smaller clusters.

Consistently two peaks are finds in the ratio of $\chi''/\chi'$. The lower temperature peak is week and corresponds to the peak in $\chi''$. The higher temperature in the ratio is strong and corresponds to the shoulder in $\chi'$ and the resistivity peak.

## DC field effect on $\chi''$:

We also discuss the dc field effect on the out of phase response. As is evident from the **Fig.6** the response is very strongly suppressed by the applied dc fields of the typical value of a few hundred Oe. This is true for the entire range of temperatures albeit for different reasons. The out of phase response is completely suppressed by fields of the order of a few hundred gauss. It may be pointed out here that the suppression of $\chi''$ is much stronger than that of $\chi'$. The peak of $\chi''$ is shifted up to 8-10 K higher temperature than the actual peak with the application of dc field up to 500 Oe in 3 and 4% Fe composition.

We understand the strong field dependence of the loss components at higher temperatures ($T>T_c$) as being due to the suppression of dynamical spin fluctuations in the transition region, as already mentioned in the context of the in-phase part. With the application of the applied dc field the fluctuations are suppressed and the dc field immobilizes part of the spins leading to a decrease in the in-phase part as well. The suppression of the peak with field probably corresponds to the removal of local correlations due to the applied field. The lower temperature peak is weak and corresponds to the peak in $\chi''$. The higher temperature in the ratio is strong and corresponds to the shoulder in $\chi'$ and the resistivity peak.

## IV. Conclusion:

Our study of transport and magnetic behavior of the *Fe* doped rare-earth manganites are consistent with previous studies in the sense of showing a general decrease of $T_c$ with increasing *Fe* content. In view of the reports on these doped materials regarding the formation of spin clusters, it seems reasonable to argue that the effect of the field is to decrease the spin disorder, both above and below $T_c$, thereby increasing the spin correlations and the conductivity. At $T_c$ where the correlations are not fully established, the field apparently helps stabilize the ferromagnetic correlations and leads to large resistivity changes in the doped materials. The low frequency dynamical response of the CMR compositions studied, particularly $La_{0.65}Ca_{0.35}Mn_{1-y}Fe_yO_3$ with $0.01<y<0.1$, is consistent with microscopic studies which indicate an inhomogeneous magnetic transition. Our results are consistent with the development of an extended cluster or region where the critical fluctuations and their related losses are strongly suppressed on the application of dc fields. We find that the pattern of the losses is consistent with the rapid growth of clusters probably up to some temperature and continued presence of these smaller clusters well below this temperature. These features are also reflected in the resistivity behavior. It is also noticeable that below $T_c$ the zero field resistivity and the ac magnetization both show the effects we may attribute to increasing spin disorder. This latter effect is consistent with increasing magnetic anisotropy. The effects of the dc field on the susceptibilities, $\chi'$ and $\chi''$, are explainable within the context of changes in the local slope of the magnetization curve and the suppression of fluctuations, away and close to the critical region, respectively. What is remarkable however is the very strong suppression of the out of phase part even for moderate dc fields ~550's of Oe. This very strong sensitivity of the out of phase response, and the very low ac fields required, suggests that the effect could be utilized to sense magnetic fields e.g. in recording devices. If the ac response of the CMR sensing head is monitored, the very rapid change in the out of phase component with field could be used as a signal. It is noticeable that the field sensitivity is strongest close to $T_c$ and hence in CMR compounds with $T_c$'s exceeding room temperatures the applicability could become more feasible. It is well known that DE mediates ferromagnetism and metallic conduction. The transport and magnetic result shown above clearly demonstrate that the partial replacement of Mn by Fe favors insulating and AF behavior, opposing the effect of double exchange (DE). Since Fe doping is the direct replacement of $Mn^{+3}$ by $Fe^{+3}$, the experimental results suggests that the sites that are now occupied by $Fe^{+3}$ can no longer effectively participate in the DE process. The mechanism that $Fe^{+3}$ terminate the DE process arises purely from the electronic structure of the materials.

**Figure Caption.**

Figure 1: Shows the dependence of resistivity as a function of temperature for the $La_{0.65}Ca_{0.35}Mn_{1-y}Fe_yO_3$ (y = 0.0 – 0.1) curve (a), (b), (c), (d), (e), (f), (g) and (h) for x=0.00, 0.01, 0.02, 0.03, 0.05, 0.07, 0.08, 0.10 respectively.

Figure 2: Temperature dependence of the real part of the susceptibility for $La_{0.65}Ca_{0.35}Mn_{1-x}Fe_xO_3$ with 0.01<x<0.1, taken ZFC (warming) curves at 5 Oe ac probing field with the superimposed dc field up to 550 Oe at a fixed frequency of 131 Hz. The symbols used in different curves for different composition are, ■, ●, ▲, ▼, ♦, +, ×, * and − for x= 0.00, 0.01, 0.02, 0.03, 0.04, 0.05, 0.07, 0.08 and 0.10 respectively.

Figure 3: The data for $\chi'$ in the Zero field cooled (ZFC) state shown the variation of the percentage change of the in-phase part of the susceptibility vs temperature. The symbols used for different Fe compositions are: ■ for x=0.01, ● for x=0.02, ▲ for x=0.04, ▼ for x=0.05, ♦ for x=0.07 and + is for x=0.08.

Figure 4: Typical temperature dependence of the out of phase part $\chi''$, with f=311Hz. and $h_{ac}$=5 Oe, the symbols used are ■ for x=0.00, ● for x=0.01, and ▲ for x=0.02, Fe doped sample shown for f=311Hz. and $h_{ac}$=5 Oe.

Figure 5: Temperature dependence of the in-phase, out-of-phase part of susceptibility and resistivity for 1% Fe composition are shown where the first peak or shoulder in $\chi''$, appears (257 K), somewhat below the maximum in resistance R, and just at the mid of the susceptibility curve. The data have been normalized to unity.

Figure 6: The dc field (H=550 Oe) effects are shown on the imaginary part of the susceptibility for 1% Fe composition. As is evident the response is very strongly suppressed by the applied dc fields of the typical value of a few hundred Oe

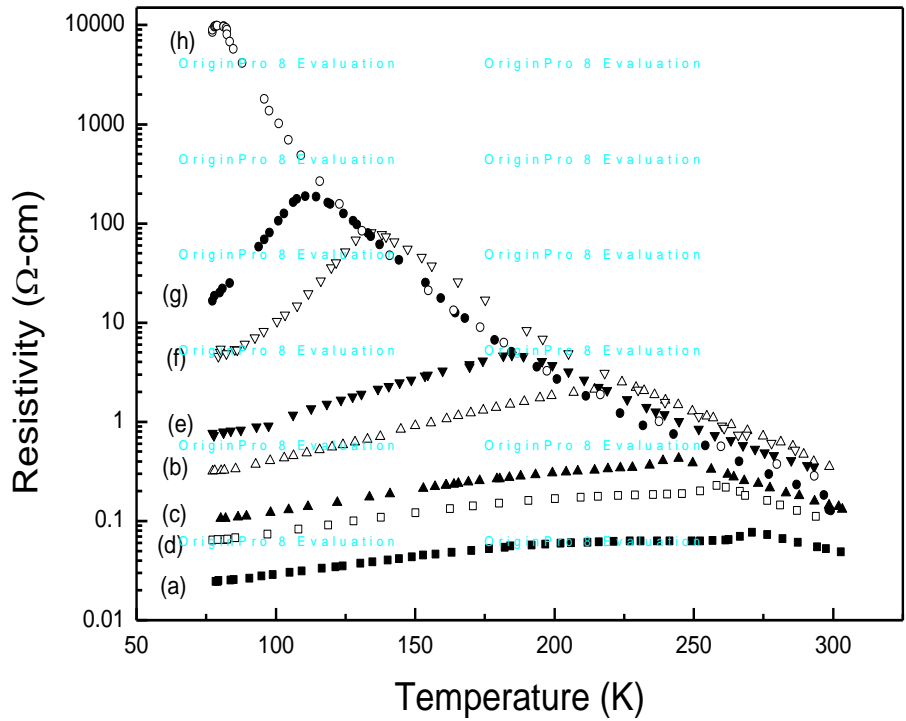

Fig. 1: Temperature dependence of the resistivity for $La_{0.65}Ca_{0.35}Mn_{1-x}Fe_xO_3$. curve (a), (b), (c), (d), (e), (f), (g) and (h) are for x=0.00, 0.01, 0.02, 0.03, 0.05, 0.07, 0.08, 0.10 respectively.

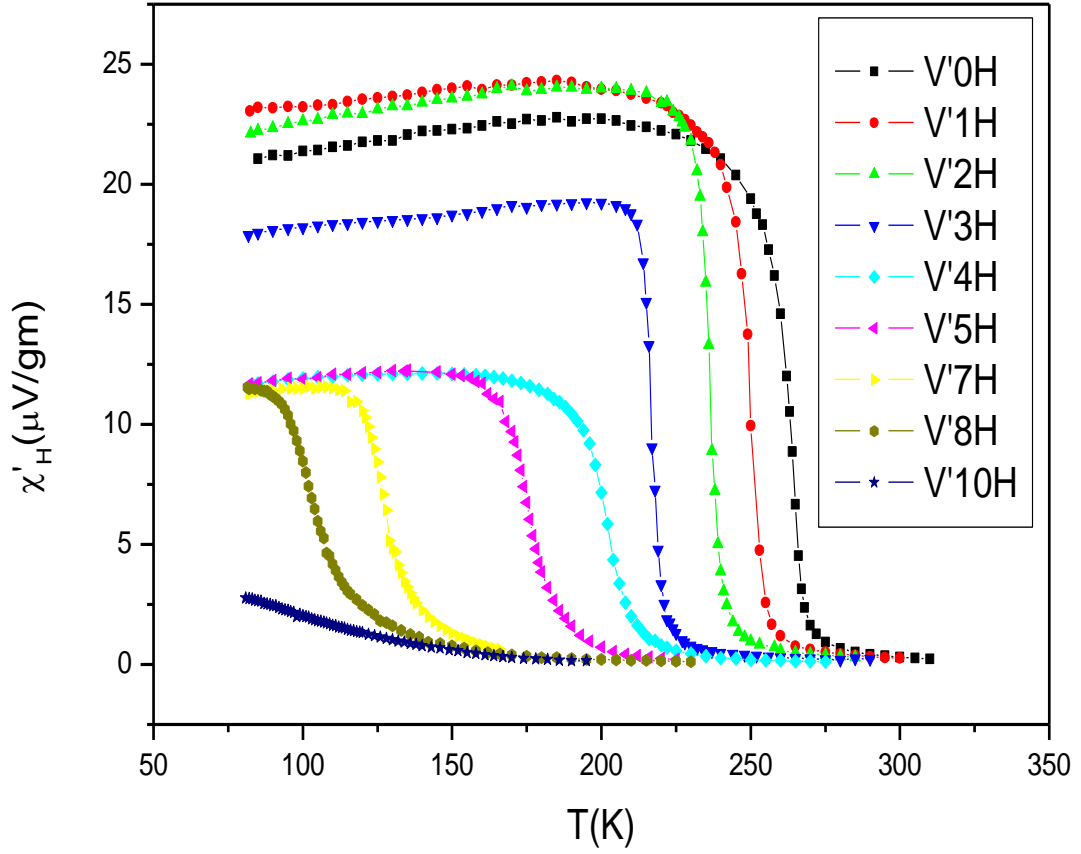

Fig. 2: Temperature dependence of the real part of the susceptibility for $La_{0.65}Ca_{0.35}Mn_{1-x}Fe_xO_3$ with $0.01<x<0.1$, taken ZFC (warming) curves at 5 Oe ac probing field with the superimposed dc field up to 550 Oe at a fixed frequency of 131 Hz. The symbols used in different curves for different composition are, ■, ●, ▲, ▼, ♦, +, ✕, ✱ and − for x= 0.00, 0.01, 0.02, 0.03, 0.04, 0.05, 0.07, 0.08 and 0.10 respectively.

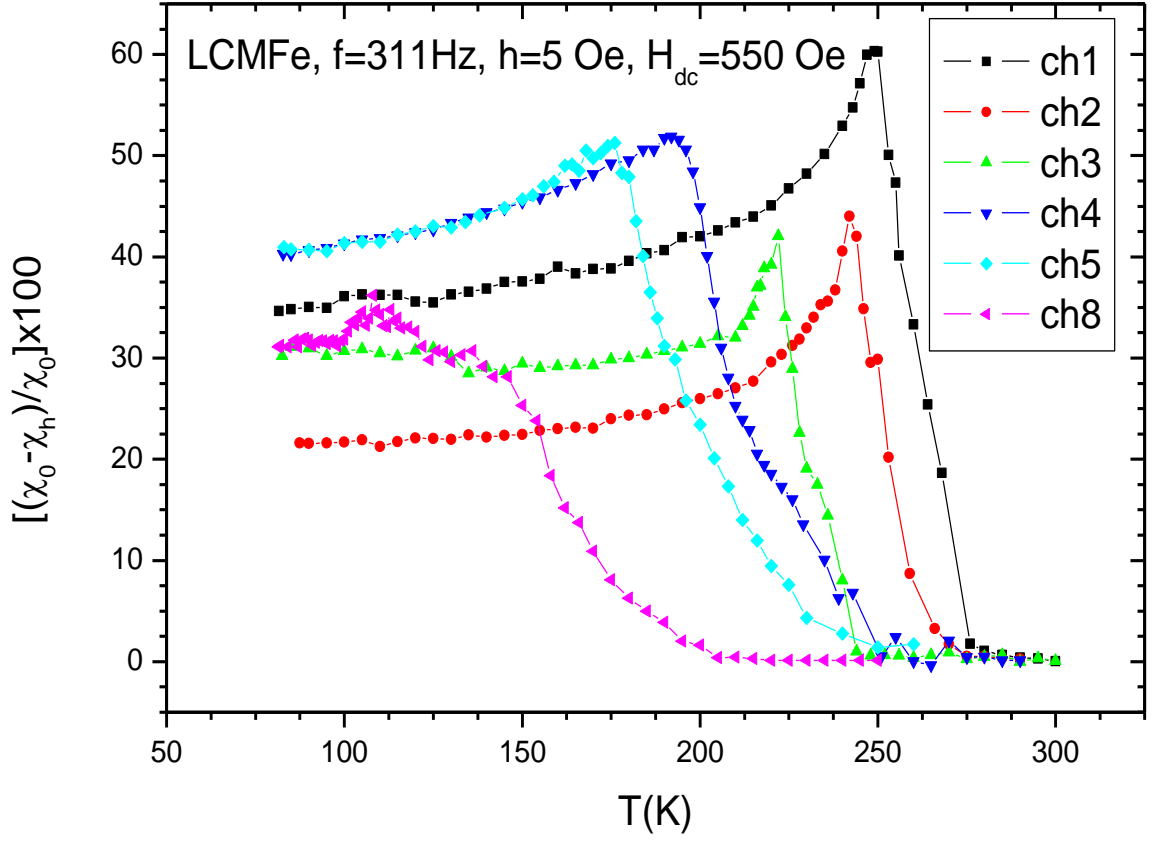

Fig. 3: The data for $\chi'$ in the Zero field cooled (ZFC) state shown the variation of the percentage change of the in-phase part of the susceptibility vs temperature. The symbols used for different Fe compositions are: ■ for x=0.01, ● for x=0.02, ▲ for x=0.04, ▼ for x=0.05, ♦ for x=0.07 and + is for x=0.08.

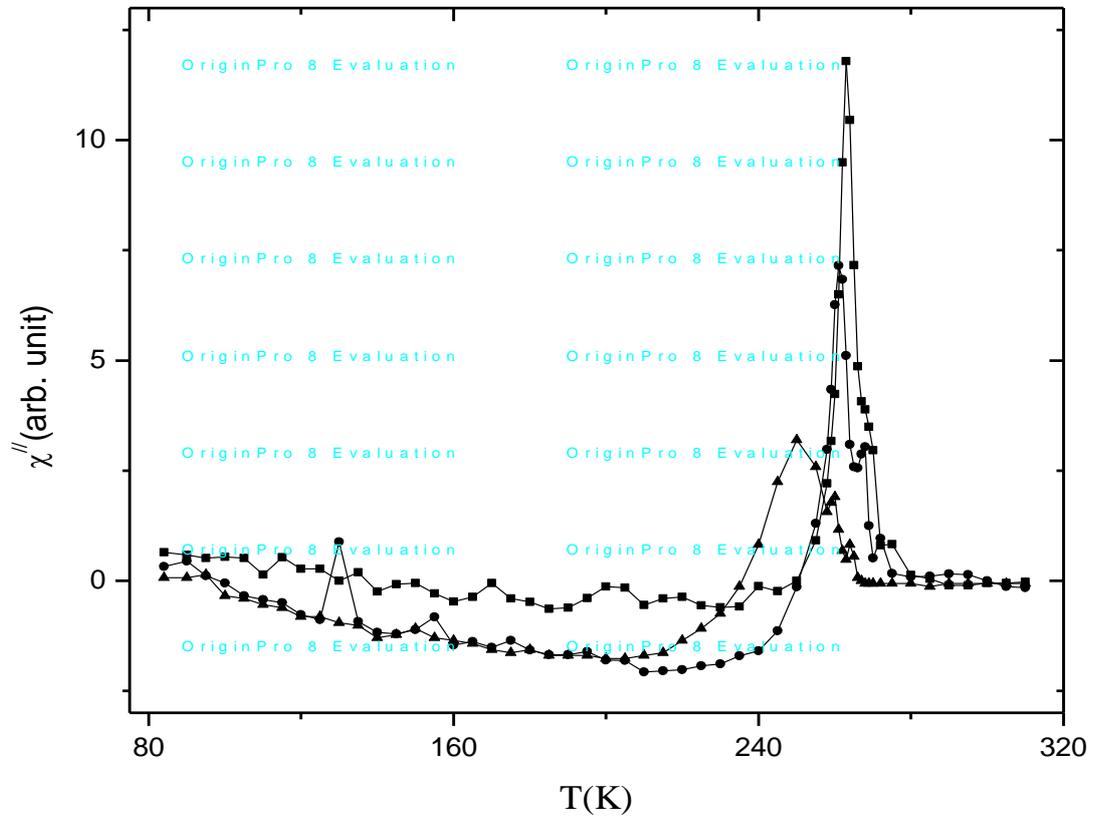

Figure 4: Typical temperature dependence of the out of phase part $\chi^{''}$, with f=311Hz. and $h_{ac}$=5 Oe, the symbols used are ■ for x=0.00, ● for x=0.01, and ▲ for x=0.02, Fe doped sample shown for f=311Hz. and $h_{ac}$=5 Oe.

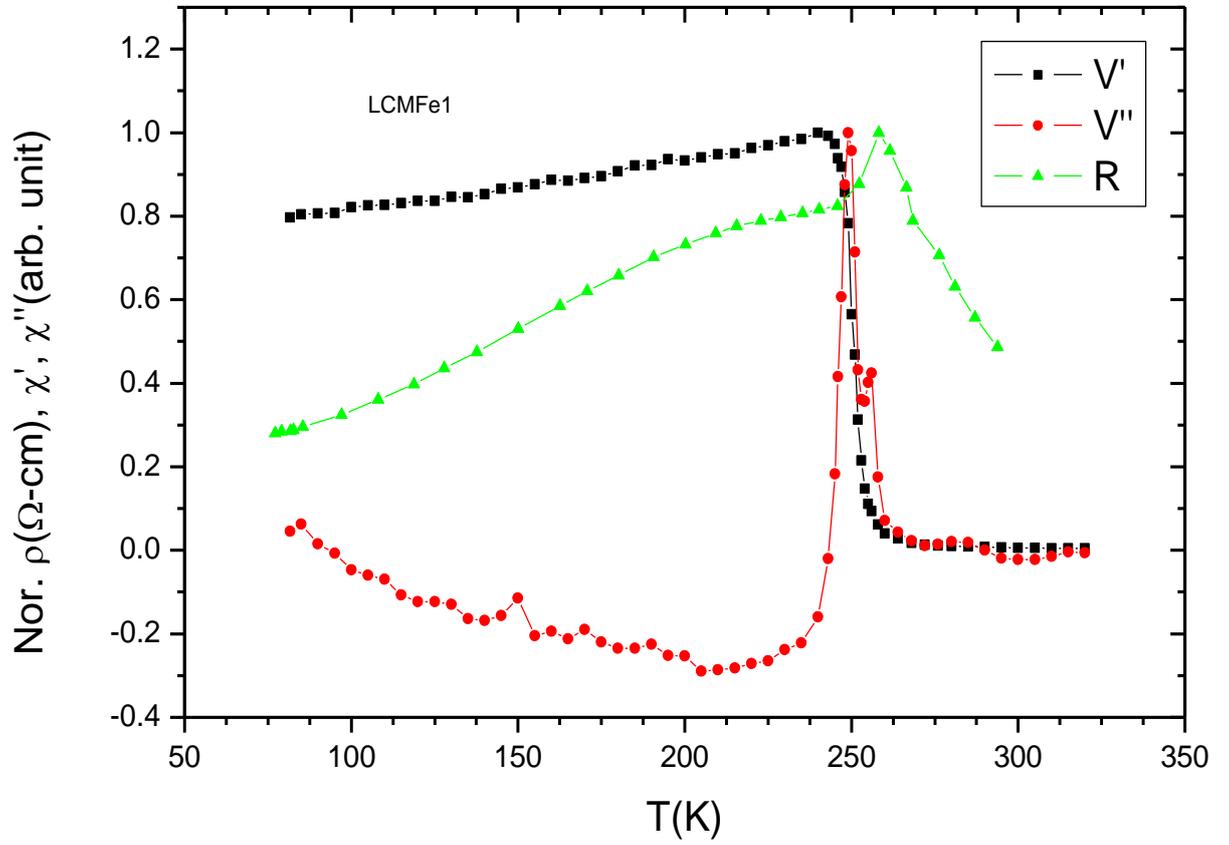

Fig.5: Temperature dependence of the in-phase, out-of-phase part of susceptibility and resistivity for 1% Fe composition are shown where the first peak or shoulder in $\chi''$, appears (257 K), somewhat below the maximum in resistance R, and just at the mid of the susceptibility curve. The data have been normalized to unity.

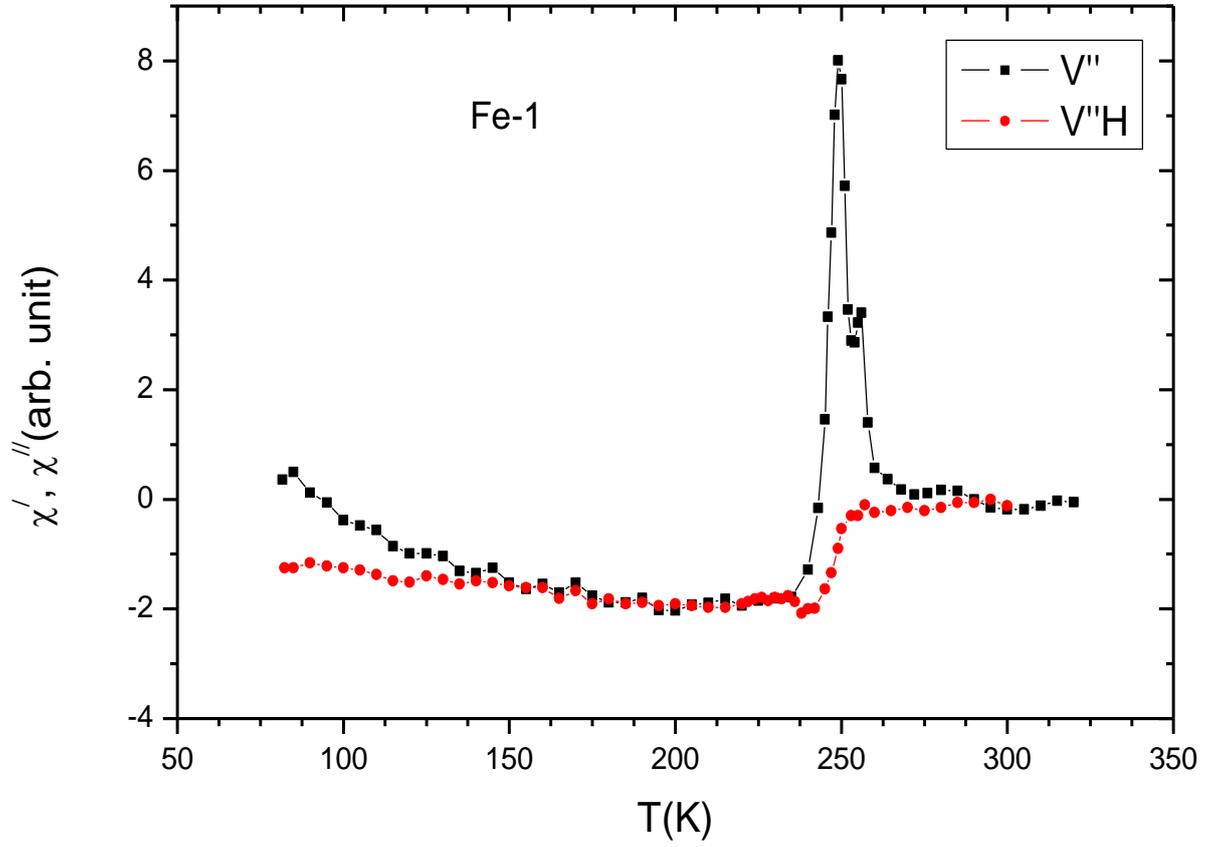

Fig. 6: The dc field (H=550 Oe) effects are shown on the imaginary part of the susceptibility for 1% Fe composition. As is evident the response is very strongly suppressed by the applied dc fields of the typical value of a few hundred Oe.